\documentclass[prx,twocolumn,amsfonts,amsmath,floatfix]{revtex4}

\usepackage{graphicx}
\usepackage{epsfig}
\usepackage{epstopdf}
\usepackage[usenames,dvipsnames]{xcolor}
\usepackage{braket}
\usepackage{hyperref}
\usepackage{amsthm}
\usepackage{enumitem}

\newcommand{\be}{\begin{equation}}
\newcommand{\ee}{\end{equation}}
\newcommand{\ba}{\begin{eqnarray}}
\newcommand{\ea}{\end{eqnarray}}

\newcommand{\bc}{\begin{center}}
\newcommand{\ec}{\end{center}}
\newcommand{\beq}{\begin{equation}}
\newcommand{\eeq}{\end{equation}}
\newcommand{\beqq}{\begin{equation*}}
\newcommand{\eeqq}{\end{equation*}}
\newcommand{\beqa}{\begin{align}}
\newcommand{\eeqa}{\end{align}}
\newcommand{\barr}{\begin{array}}
\newcommand{\earr}{\end{array}}
\newcommand{\bi}{\begin{itemize}}
\newcommand{\ei}{\end{itemize}}

\newtheorem{theo}{Theorem}

\newtheorem{coro}{Corollary}
\newtheorem{defi}{Definition}

\begin{document}

\title{Optimal quantum-programmable projective measurement with linear optics}

\author{Ulysse Chabaud$^{1}$}
\email{ulysse.chabaud@gmail.com}
\author{Eleni Diamanti$^1$}
\author{Damian Markham$^1$}
\author{Elham Kashefi$^{1,2}$}
\author{Antoine Joux$^3$}
\email{antoine.joux@m4x.org}
\address{$^1$ Laboratoire d'Informatique de Paris 6, CNRS, Sorbonne Universit\'e, 4 place Jussieu, 75005 Paris}
\address{$^2$ School of Informatics, University of Edinburgh, 10 Crichton Street, Edinburgh, EH8 9AB}
\address{$^3$ Chaire de Cryptologie de la Fondation SU, Sorbonne Universit\'e, Institut de Math\'ematiques de
Jussieu -- Paris Rive Gauche, CNRS, INRIA, Universit\'e Paris Diderot, Campus Pierre et Marie Curie, 4 place Jussieu, 75005 Paris}
\date{\today}


\begin{abstract}

We present a scheme for a universal device which can be programmed by quantum states to approximate a chosen projective measurement to a given precision. Our scheme can be viewed as an extension of the swap test to the instance where one state is supplied many times. As such, it has many potential applications given the variety of quantum information tasks which make use of the swap test.
In particular, we show that our scheme is optimal for state discrimination under the one-sided error requirement, and optimally approximates any projective measurement.
Furthermore, we propose a practical implementation of our scheme with passive linear optics, which involves a simple interferometer composed only of balanced beam splitters.

\end{abstract}


\maketitle


\section{Introduction}

In a typical experiment performing a quantum measurement, the choice of measurement is encoded in macroscopic, classical, information in the experimental set up.
For example it can be encoded into the reflectivity of a beam splitter, the phase in the branch of an interferometer or the spacial direction of a Stern Gerlach device. 
Often these choices are made beforehand and fixed. In some cases they can be programmed in a single set up (for example using thermo-optic phase shifters \cite{carolan2015universal}).
In all these cases, however, the choice of measurement basis is effectively programmed classically. 

In this work we consider the case where the choice of measurement is instead controlled by a quantum state.
There are several reasons why one may consider a quantum state to control the choice of measurement. 
This state may be an output of a quantum computer, or a communication protocol, for example, which is not known before hand and only accessible as a quantum state. For example, in the cryptographic setting, non-orthogonal states can be used to remotely program a measurement which allows one to test the behaviour of a remote party. This is the essence behind the delegated blind verified quantum computation in~\cite{fitzsimons2017unconditionally}. 
At a fundamental level quantum programmable measurements separate as much as possible the choice of measurement basis and the bulk of the physical measurement apparatus, which could be interesting in probing
foundational questions, for example in tests of contextuality where  information about which measurements are being carried out leads to loopholes \cite{meyer1999finite,clifton2000simulating,winter2014does}.

 A related and, in a sense, more general problem is that of a programmable quantum computer, where a quantum program state is used to encode a unitary to be run on a generic quantum computing device (gate array), first proposed by Nielsen and Chuang \cite{nielsen1997programmable}. 
 There it was shown that to do so deterministically requires orthogonal program states for every different unitary. To use the continuous parameters available in quantum states to encode more computations, the best one can do is probabilistic.
In principle these techniques can be used to program quantum measurements.
Indeed since the original proposal there have been several alternative schemes, extensions and applications, including programmable quantum state discriminators and measurements \cite{vidal2000storage,duvsek2002quantum,rovsko2003generalized,ziman2005realization,bergou2006programmable}.
These results, however, are either too general to consider the type of efficiency we show here, or specialized to tasks which are different from our simple setting (for example state discrimination \cite{bergou2006programmable}).

\begin{figure}
	\begin{center}
		\includegraphics[width=3.1in]{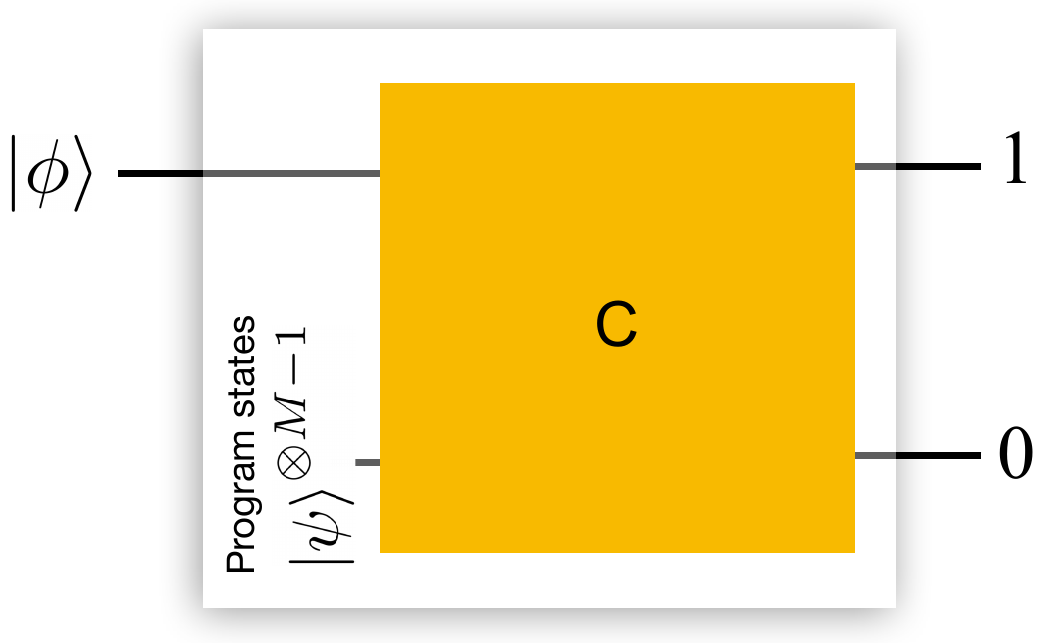}
		\caption{Programmable projective measurement. Given an input $\ket{\phi}$ and $M-1$ program registers $|\psi\rangle^{\otimes M-1}$, and allowing for possible ancillas (not pictured here), we apply some circuit $C$, independent of $|\psi\rangle$, and output a binary result where $0$ is associated to projecting onto $|\psi\rangle$ and $1$ to its complement.}
		\label{fig:programmable projective measurement definition}
	\end{center}
\end{figure}

We cast our problem as follows, illustrated in Fig.~\ref{fig:programmable projective measurement definition}. 
One has $M-1$ program registers each prepared in the state $|\psi\rangle$ corresponding to the choice of measurement basis, and a single input register prepared in some state $\ket{\phi}$. 
Our aim is to output a classical bit corresponding to a projective measurement, where $0$ represents the outcome $|\psi\rangle$ and $1$ represents its complement.
In an ideal measurement the result $0$ would occur with probability $|\langle  \phi |\psi\rangle|^2$. 
However, this is impossible for finite $M$. 
This follows from standard arguments based on the linearity of quantum mechanics, in analogy to necessity of orthogonal program states for computation mentioned above. (See for example \cite{nielsen1997programmable} for the case of programmable universal quantum computation, which easily extends to our case).
We can thus only ever approximate perfect measurements.  In our case we parametrise this approximation by $\epsilon$, requiring that the result $0$ is returned with probability $\epsilon$-close to $|\langle  \phi |\psi\rangle|^2$ (see section \ref{sec:circuit} for a formal definition).

We present a scheme which achieves this optimally in terms of how $\epsilon$ scales with $M$, under the condition that if the input is $\ket{\psi}$, the measurement always returns $0$. 
This so-called one-sided error requirement~\cite{buhrman1999one} makes sense for various potential applications where it is important not to be wrong for this answer. One such example is the link between our scheme and the swap test~\cite{buhrman2001quantum}. 

In the swap test, two unknown quantum states are compared using a controlled-swap operation. This test is especially relevant for the task of state discrimination. 
The general task of assessing if a set of $M$ arbitrary states are identical has been addressed in~\cite{chefles2004unambiguous,kada2008efficiency}. 
To solve this in generality requires controlled permutations for all possible permutations and therefore scales exponentially in circuit size.
If one restricts oneself to the case where one has $M/2$ copies of one state and $M/2$ copies of the other, one can apply the construction in~\cite{kada2008efficiency} to get an optimal result.
However, this scaling is not much better than simply doing the original swap test $M/2$ times, yet the corresponding test is much more difficult.

From this point of view, the interesting cases of two states comparison is if one has an asymmetric number of one compared state compared to the other. In the most extreme case one would have just one copy of one state and $M-1$ copies of the other, which is exactly the case we consider for our programmable projective measurement, viewing the program state as the one we have many copies of.
In particular, the $M=2$ case reduces to the swap test.

Moreover, the swap test has been shown equivalent to the linear optical Hong-Ou-Mandel effect~\cite{garcia2013swap}. Generalising this equivalence, we present a practical solution to our problem with linear optics, using the Hadamard interferometer~\cite{crespi2015suppression,crespi2016suppression}. 

The next sections are organised as follows. In Sec.~\ref{sec:swap} we introduce the circuits for the swap test and its generalisation, the swap test of order $M$. 
We show in Sec.~\ref{sec:circuit} that these circuits can be used for programmable projective measurement and prove their optimality.
We then present in Sec.~\ref{sec:optic} a simple linear optical interferometer to implement our scheme. 
For completeness, we introduce in Sec.~\ref{sec:general} a general family of interferometers which reproduce the appropriate statistics.
We conclude with an interpretation of our results and discuss various applications in Sec.~\ref{sec:concl}.


\section{Swap circuit of order $M$}
\label{sec:swap}

The swap test~\cite{buhrman2001quantum} provides an efficient probabilistic tool to compare two unknown quantum states. It takes as input two quantum states $\ket{\phi}$ and $\ket{\psi}$ that are not entangled and outputs $0$ with probability $\frac{1}{2}+\frac{1}{2}|\braket{\phi|\psi}|^2$ and $1$ with probability $\frac{1}{2}-\frac{1}{2}|\braket{\phi|\psi}|^2$, where $\braket{\phi|\psi}$ is the overlap between the states $\ket{\phi}$ and $\ket{\psi}$. When the measurement outcome is $0$ (resp. $1$), we conclude that the states were identical (resp. different), up to a global phase. 

\begin{figure}
	\begin{center}
		\includegraphics[width=2.5in]{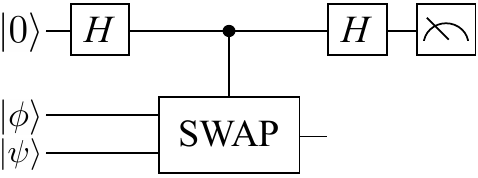}
		\caption{Circuit representation of a swap test. The ancilla qubit is measured in the computational basis.}
		\label{fig:SWAP}
	\end{center}
\end{figure}

A circuit implementing the swap test is represented in Fig.~\ref{fig:SWAP}, where an ancilla is first prepared in the $|+\rangle$ state by a Hadamard gate
\be
H=\frac{1}{\sqrt2}\begin{pmatrix} 1 & 1 \\ 1 & -1 \end{pmatrix},
\label{matrixH}
\ee
which controls a swap between the two systems being tested. 

The swap test meets the so-called one-sided error requirement~\cite{buhrman1999one}, i.e. if the input states are identical, the test will always declare them as identical. On the other hand, if the input states are different, the test can obtain a wrong conclusion and declare the states identical. The probability that this happens is strictly less than $1$, hence by repeating the test various times, the probability that the sequence of tests never answers $1$ can be brought down arbitrarily close to zero, exponentially fast. However, the swap test is destructive, in the sense that the output states of a previous test cannot be reused for a new test because they become maximally entangled during the test~\cite{garcia2013swap}. This means that in order to boost the correctness of the test in this manner, multiple copies of both states must be available. 

Let $M\ge2$. We introduce the following generalisation of the swap test, in the context where one has access to various copies of a reference state $\ket{\psi}$ but to only a single copy of the other tested state $\ket{\phi}$:

\begin{defi} The swap test of order $M$ is a binary test that takes as input a state $\ket{\phi}$ and $M-1$ copies of a state $\ket{\psi}$, and outputs $0$ with probability $\frac{1}{M}+\frac{M-1}{M}|\braket{\phi|\psi}|^2$ and $1$ with probability $(\frac{M-1}{M})(1-|\braket{\phi|\psi}|^2)$. If the outcome $0$ (resp. $1$) is obtained, the test concludes that the states $\ket{\phi}$ and $\ket{\psi}$ were identical (resp. different).
\label{defswap}
\end{defi}

Such a test clearly satisfies the one-sided error requirement.
\begin{figure}
\begin{center}
\includegraphics[width=3.1in]{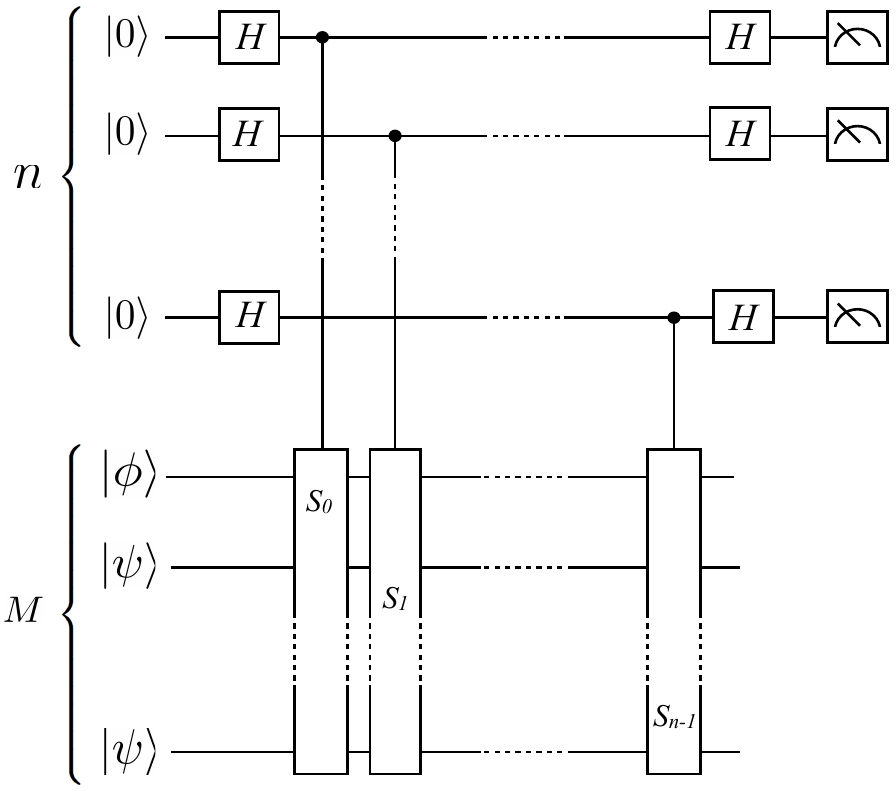}
\caption{Swap circuit of order $M$. The unitaries $S_k$ are tensor products of swap gates described in the main text~(\ref{Sk}). The $n=\log M$ ancilla qubits are measured in the computational basis at the end of the computation. The probability of obtaining $0$ for all measurement outcomes is $\frac{1}{M}+\frac{M-1}{M}|\braket{\phi|\psi}|^2$.}
\label{fig:swap_circuit}
\end{center}
\end{figure}
In the following, we restrict to the swap test of order $M$ when $M$ is a power of $2$, writing $n=\log M$. We introduce the swap circuit of order $M$ (Fig.~\ref{fig:swap_circuit}), that acts on $M$ input qubits by applying $n$ consecutive layers of products of swap gates controlled by $n$ ancilla qubits. These ancilla qubits are first initialised in the $\ket{+}$ state using Hadamard gates. Then, they are used as control qubits for the gates $S_0,\dots,S_{n-1}$, which can be applied in any order, where for all \mbox{$k\in\{0,\dots,n-1\}$}
\begin{equation}
S_k=\underset {\substack{ i\in \left[ 0,{ 2 }^{ k }-1 \right] ,\\j\in \left[ 0,{ 2 }^{ n-k-1 }-1 \right] } }{ \bigotimes} \text{SWAP}\left[j { 2 }^{ k+1 }+i , j { 2 }^{ k+1 }+i+{ 2 }^{ k }\right] ,
\label{Sk}
\end{equation}
with SWAP$[i,j]$ being the unitary operation that swaps the $i^{th}$ and $j^{th}$ qubits for $i,j\in\{0,\dots,M-1\}$. These controlled gates are applied to the input states $\ket{\phi},\ket{\psi},\dots,\ket{\psi}$ (one copy of a state $\ket{\phi}$ and $M-1$ copies of a state $\ket{\psi}$). Finally, a Hadamard gate is applied to each ancilla, which is then measured in the computational basis.
By a simple induction, we obtain that the probability of obtaining the outcome $0$ for all ancilla qubits is the squared norm of the following state:
\begin{equation}
\frac{1}{M}(\ket{\phi\psi\dots\psi}+\ket{\psi\phi\dots\psi}+\dots+\ket{\psi\dots\psi\phi}),
\label{perror}
\end{equation}
which only depends on the overlap between the states $\ket{\phi}$ and $\ket{\psi}$. More precisely,
\begin{equation}
\text{Pr}(0,\dots,0)=\frac{1}{M}+\frac{M-1}{M}|\braket{\phi|\psi}|^2.
\label{sswap}
\end{equation}
The swap circuit of order $M$ thus implements the swap test of order $M$. Indeed, if the outcome $(0,\dots,0)$ is obtained, the test outputs $0$ and we conclude that the states were identical, while for any other outcome the test outputs $1$ and we conclude that the states were different. Note that in the case where $M=2$, the scheme reduces to the original swap test.

\begin{figure}
\begin{center}
\includegraphics[width=3.45in]{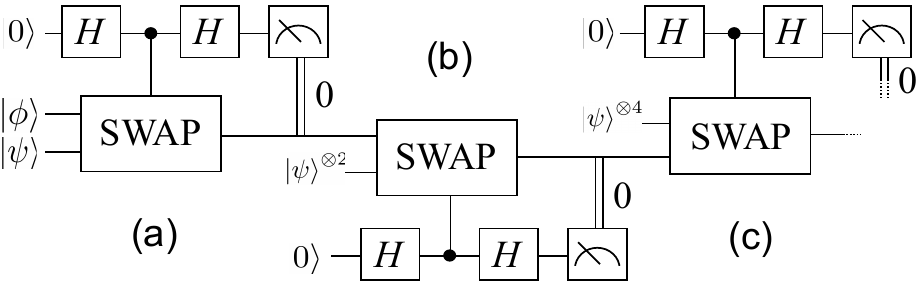}
\caption{The simplified swap circuit of order $M$ consisting in $n=\log M$ consecutive swap tests. (a) The first swap test compares the input states $\ket{\phi}$ and $\ket{\psi}$. (b) If this test is not able to tell apart the input states, i.e. if its outcome is $0$, then the second swap test compares the bipartite output state of the first test with the state $\ket{\psi}^{\otimes2}$. (c) If this test outcome is again $0$, then the third swap test compares the quadripartite output state of the second test with the state $\ket{\psi}^{\otimes4}$, and so on. If the $n$ outcomes are $0$, the test concludes that the states $\ket{\phi}$ and $\ket{\psi}$ were identical.}
\label{fig:swap_circuit_2}
\end{center}
\end{figure}

Because the $M-1$ last input states are identical, swapping them acts as the identity. This can be used to simplify the swap circuit of order $M$ by replacing the $n=\log M$ layers of swap gates in Eq.~(\ref{Sk}) by the following $n$ layers $S_0',\dots,S_{n-1}'$, which have to be applied in this order:
\begin{equation}
S_k'=\overset { { 2 }^{ k }-1 }{ \underset { l=0 }{ \bigotimes  }  } \text{SWAP}\left[ l,l+2^k \right].
\label{Sk2}
\end{equation}
This reduces the total number of swap gates from $\frac{M\log M}{2}$ to $M-1$ without changing the number of ancilla qubits. This circuit has a simple structure of $n=\log M$ consecutive swap tests (Fig.~\ref{fig:swap_circuit_2}). 
For $k\in\{0,\dots,n-1\}$, conditioned on all the previous outputs being $0$, the $k^{th}$ swap test compares the output state of the previous test and the state $\ket{\psi}^{\otimes2^{k}}$. Here, the swap test of two multipartite quantum states consists in applying a swap test to each of their corresponding subsystems. However, this multipartite swap test uses only a single ancilla qubit controlling the product of swap gates, as in Eq.~(\ref{Sk2}), instead of an ancilla qubit for each pair of subsystems.

We now prove the optimality of the swap test of order $M$ under the one-sided error requirement, i.e. we show that it achieves the lowest error probability in comparing states $|\phi\rangle$ and $\ket{\psi}$ given $M-1$ copies of $|\phi\rangle$ and one copy of $\ket{\psi}$ such that the one-sided error requirement is satisfied.

For this purpose, we first derive a more general result. In Ref.~\cite{kada2008efficiency}, the authors consider the problem of testing if $M$ quantum states are identical or not (the so called \textit{identity test}), with the promise that all the states are pairwise identical or orthogonal. 
In particular, they show that the optimal value for the error probability of any identity test with these assumptions satisfying the one-sided error requirement is $\frac{1}{M}$. We extend this result to the case where the states to be compared are no longer assumed pairwise identical or orthogonal:

\begin{theo} Under the one-sided error requirement, any identity test of $M$ unknown quantum states $\ket{\psi_0},\dots,\ket{\psi_{M-1}}$ has an error probability at least
\be
\frac{1}{M!}\sum_{\sigma\in\mathcal{S}_M}{\prod_{k=0}^{M-1}{\braket{\psi_k|\psi_{\sigma(k)}}}},
\ee
where $\mathcal{S}_M$ is the symmetric group over $\{0,\dots,M-1\}$.
\label{th:sopt}
\end{theo}

\textbf{Proof.} An identity test satisfying the one-sided error requirement can only be wrong when declaring identical states (outputting $0$) that were not identical. Hence, to prove Theorem~\ref{th:sopt}, it suffices to lower bound the probability of outputting $0$ for any identity test. This is done by showing that the optimal identity test consists in a projection onto the symmetric subspace of the input states Hilbert space. We give a detailed proof in the Appendix~\ref{app:opti}. 

\qed

Applying Theorem~\ref{th:sopt} with \mbox{$\ket{\psi_0\dots\psi_{M-1}}=\ket{\phi\psi\dots\psi}$} implies that the value $\frac{1}{M}+\frac{M-1}{M}|\braket{\phi|\psi}|^2$ is a lower bound for the error probability of any identity test of $M$ states $\ket{\phi},\ket{\psi},\dots,\ket{\psi}$ (one copy of a state $\ket{\phi}$ and $M-1$ copies of a state $\ket{\psi}$). With Definition~\ref{defswap} we directly obtain the following result:

\begin{coro} The swap test of order $M$ has optimal error probability $\frac{1}{M}+\frac{M-1}{M}|\braket{\phi|\psi}|^2$ under the one-sided error requirement.
\label{co:0}
\end{coro}

The swap circuit of order $M$ is thus optimal for quantum state identity testing with an input $\ket{\phi},\ket{\psi},\dots,\ket{\psi}$, under the one-sided error requirement, since it implements the swap test of order $M$. In the next section, we show that the swap circuit of order $M$ can be used to implement a programmable projective measurement.


\section{Circuit for programmable projective measurement}
\label{sec:circuit}

Given that a projective measurement with respect to a state $\ket{\psi}$ is a process that takes as input a state $\ket{\phi}$ and outputs $0$ with probability $|\braket{\phi|\psi}|^2$ and $1$ with probability $1-|\braket{\phi|\psi}|^2$, we introduce the natural notion of projective measurement with finite error:

\begin{defi} Given a quantum state $\ket{\psi}$ and $\epsilon>0$, a projective measurement with error $\epsilon$ with respect to the reference state $\ket{\psi}$ is a process that takes as input a quantum state $\ket{\phi}$ and outputs $0$ with probability $P(0)$ and $1$ with probability $P(1)$, such that $|P(0)-(|\braket{\phi|\psi}|^2)|\le\epsilon$ and $|P(1)-(1-|\braket{\phi|\psi}|^2)|\le\epsilon$.
\label{def:proj}
\end{defi}

Note that the two conditions in the previous definition are equivalent, since $P(0)+P(1)=1$. It will thus suffice to consider e.g. the first condition. In this context, under the one-sided error requirement, a projective measurement with any error $\epsilon$ always outputs $0$ if the input state is equal to the reference state.

\begin{theo} A swap circuit of order $M$ can be used to perform a projective measurement with error $\frac{1}{M}$ under the one-sided error requirement. Moreover, it is optimal in the sense that it uses the minimum number of copies of the reference state for achieving such an error.
\label{th:circuit}
\end{theo}

\begin{figure}
\begin{center}
\includegraphics[width=3.3in]{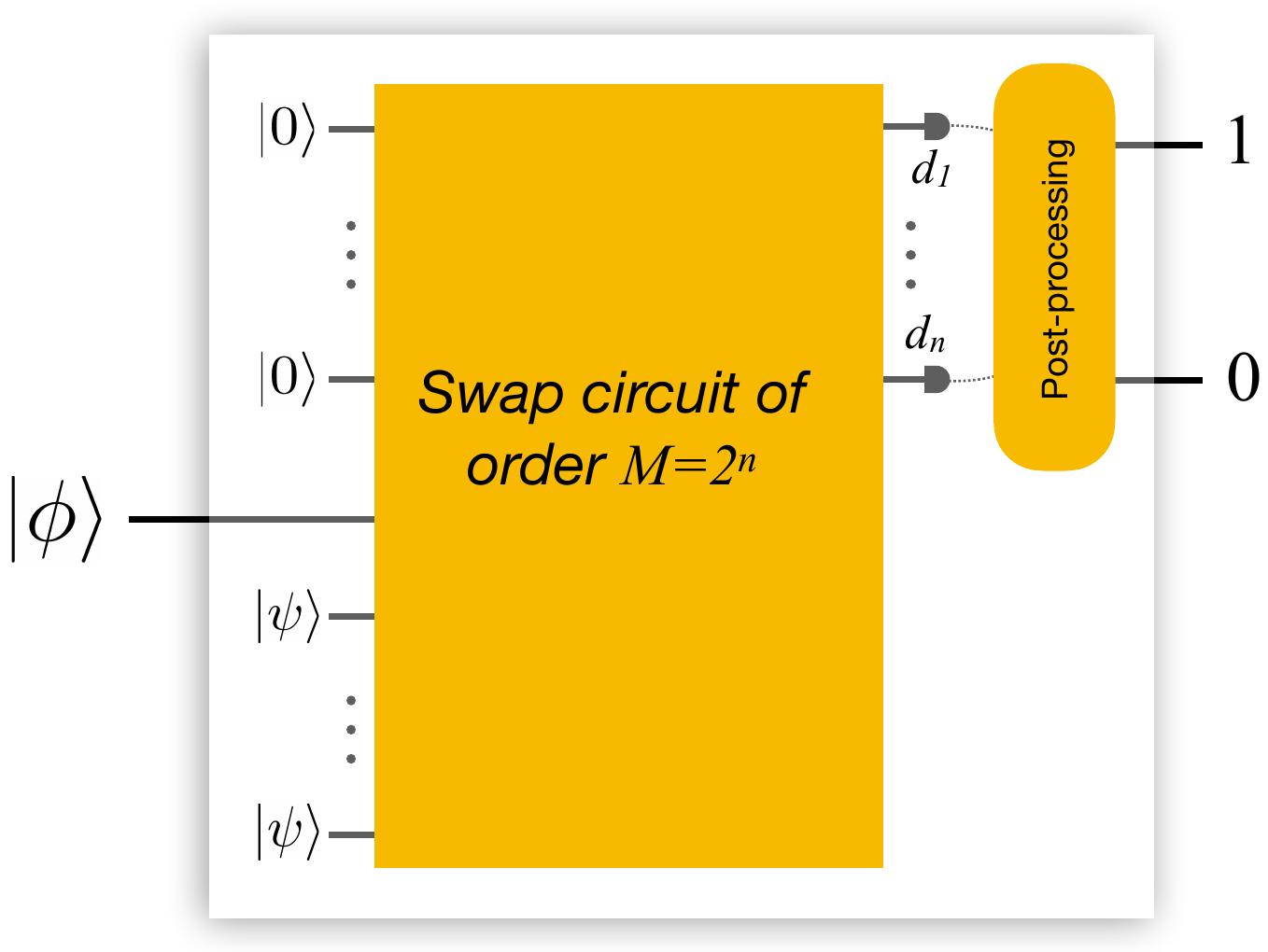}
\caption{The swap circuit of order $M$ used as a programmable projective measurement device. It takes as input a state $\ket{\phi}$ and the internal measurement outcomes are post-processed such that the device outputs $0$ with probability $\frac{1}{M}+\frac{M-1}{M}|\braket{\phi|\psi}|^2$ and $1$ with probability $\frac{M-1}{M}(1-|\braket{\phi|\psi}|^2)$. The programmable resource is the state $\ket{\psi}$ and the process uses $M-1$ copies of this state as well as $n=\log M$ ancillas.}
\label{fig:setupc}
\end{center}
\end{figure}

\textbf{Proof.} For the swap circuit of order $M$, we have Pr$(0,\dots,0)=\frac{1}{M}+\frac{M-1}{M}|\braket{\phi|\psi}|^2$, so we can consider the whole circuit except the state $\ket{\phi}$ as a black box in Fig.~\ref{fig:swap_circuit}, and post-process the measurement outcomes $D$ as follows: if $D=(0,\dots,0)$, output $0$, and output $1$ otherwise (Fig.~\ref{fig:setupc}). The setup now takes a single state $\ket{\phi}$ in input and outputs $0$ with probability $P(0)=\frac{1}{M}+\frac{M-1}{M}|\braket{\phi|\psi}|^2$, and $1$ with probability $P(1)=1-P(0)$. We have $|P(0)-(|\braket{\phi|\psi}|^2)|\le\frac{1}{M}$ and when $\ket{\phi}=\ket{\psi}$, we have $P(0)=1=|\braket{\phi|\psi}|^2$, hence this device performs a projective measurement with error $\frac{1}{M}$ and meets the one-sided error requirement. 

We now prove the optimality of this device in terms of resources, i.e. we show that any device implementing a projective measurement with error $\frac{1}{M}$ and meeting the one-sided error requirement cannot use less than \mbox{$M-1$} copies of the reference state. 

We consider a device that implements a projective measurement with error $\epsilon$, with respect to a reference state $\ket{\psi}$, using $N$ copies of this reference state. This device takes as input a quantum state $\ket{\phi}$ and outputs $0$ with probability $P_\phi(0)$ and $1$ with probability $P_\phi(1)=1-P_\phi(0)$. By Definition~\ref{def:proj}, the probability of outputting $0$ satisfies $|P_\phi(0)-(|\braket{\phi|\psi}|^2)|\le\epsilon$. When the input state $\ket{\phi}$ is orthogonal to the reference state $\ket{\psi}$, the probability $P_{\phi,\bot}(0)$ of outputting $0$ thus satisfies
\be
P_{\phi,\bot}(0)\le\epsilon. 
\label{Pbot1}
\ee
On the other hand, we can use this device to perform an identity test of $N+1$ states $\ket{\phi},\ket{\psi},\dots,\ket{\psi}$ (one copy of the state $\ket{\phi}$ and $N$ copies of the state $\ket{\psi}$): if the output $0$ (resp. $1$) is obtained we conclude that the states were identical (resp. different). This device meets the one-sided error requirement, so by Theorem~\ref{th:sopt} it has error probability at least $\frac{1}{N+1}+\frac{N}{N+1}|\braket{\phi|\psi}|^2$. This error probability corresponds to the probability of outputting $0$ when the input states are different. In particular, when the input state $\ket{\phi}$ is orthogonal to the reference state $\ket{\psi}$, the probability $P_{\phi,\bot}(0)$ of outputting $0$ thus satisfies
\be
P_{\phi,\bot}(0)\ge\frac{1}{N+1}. 
\label{Pbot2}
\ee
Combining both inequalities~(\ref{Pbot1},\ref{Pbot2}) we obtain $\frac{1}{N+1}\le\epsilon$ or equivalently $N\ge\frac{1}{\epsilon}-1$. For $\epsilon=\frac{1}{M}$, this amounts to $N\ge M-1$, which completes the proof.

\qed\\

Theorem~\ref{th:circuit} implies that given a large enough swap circuit and the ability to produce many copies of a state $\ket{\psi}$, one can projectively measure any state with respect to the state $\ket{\psi}$ up to arbitrary small error. This error scales as the inverse of the number of copies. The circuit can thus be used as a programmable projective measurement device, where the programmable resource is the reference state $\ket{\psi}$ whose number of copies can be adjusted to control the precision of the measurement (Fig.~\ref{fig:setupc}).

The implementation of the swap circuit of order $M$ is however challenging, due to the presence of many controlled-swap gates. In order to lower the implementation requirements, we study in the next section the Hadamard interferometer and show that its statistics can be efficiently post-processed to reproduce those of a swap circuit of order $M$, without the need for ancillas. This comes at the cost that the device no longer has a quantum output, which however does not matter for most applications. In particular we show that the Hadamard interferometer provides a simple linear optical platform for implementing the programmable projective measurement that we have described.


\section{Interferometer for programmable projective measurement}
\label{sec:optic}

In what follows, we consider optical unitary interferometers of size $M$ which take as input one single photon in a quantum state $\ket{\phi}$ and $M-1$ indistinguishable single photons in a state $\ket{\psi}$, one in each spatial mode (the spatial modes of the interferometers are indexed from $0$ to $M-1$). These states should be thought of as encoded in additional degrees of freedom of the photons (e.g. polarisation, time-bins). The output modes are measured using photon number resolving detection. 

There exist complex amplitudes $\alpha$ and $\beta$ and a state $\ket{\psi^\bot}$ with $\braket{\psi|\psi^\bot}=0$ such that
\begin{equation}
\ket\phi=\alpha\ket\psi+\beta\ket{\psi^\bot},
\end{equation}
where $\alpha=\braket{\psi|\phi}$ and $|\alpha|^2+|\beta|^2=1$. We have the following homomorphism property for single photon states:
\begin{equation}
\ket{1_\phi}=\ket{1_{\alpha\psi+\beta\psi^\bot}}=\alpha\ket{1_\psi}+\beta\ket{1_{\psi^\bot}},
\end{equation}
where for any state $\ket{\chi}$, $\ket{1_\chi}$ is the state of a single photon encoding the state $\ket{\chi}$. It thus suffices to compute the output statistics separately when $\ket{\phi}=\ket{\psi}$ (\textit{indistinguishable case}) and when $\ket{\phi}=\ket{\psi^\bot}$ (\textit{distinguishable case}) to obtain the output statistics in the general case by linearity. The probability of detecting the photon number pattern $D=(d_0,\dots,d_{M-1})$, or equivalently that the $k^{th}$ detector detects $d_k$ photons for all $k\in\{0,\dots,M-1\}$, is then 
\begin{equation}
\begin{aligned}
\text{Pr}(D)&=|\alpha|^2\text{Pr}_i(D)+|\beta|^2\text{Pr}_d(D) \\
&=\text{Pr}_d(D)+|\braket{\phi|\psi}|^2\left[\text{Pr}_i(D)-\text{Pr}_d(D)\right],
\end{aligned}
\label{partialud}
\end{equation}
where $\text{Pr}_i(D)$ is the probability in the indistinguishable case and $\text{Pr}_d(D)$ is the probability in the distinguishable case. The single photon encoding maps identity of quantum states to distinguishability of single photons. Note that for any measurement outcome $D=(d_0,\dots,d_{M-1})$, we have $d_0+\dots+d_{M-1}=M$ since an interferometer is a passive device that does not change the total number of photons. For any interferometer of size $M$, we prove in Appendix~\ref{app:stat} the following inequality:

\begin{equation}
\text{Pr}_d(D)\ge\frac{\text{Pr}_i(D)}{M},
\label{general}
\end{equation}
for any detection pattern $D$. Combining this inequality with Eq.~(\ref{partialud}) yields
\begin{equation}
\text{Pr}(D)\ge \left(\frac{1}{M}+\frac{M-1}{M}|\braket{\phi|\psi}|^2\right)\text{Pr}_i(D).
\label{boundpr}
\end{equation}
This last expression is valid for any interferometer and can be used it to retrieve, in the context of linear optics, the error probability bound for state identity testing under the one-sided error requirement obtained in Corollary~\ref{co:0}. Indeed, assume that $E$ is a detection event, which could be a disjoint union of detection events, used for an identity test: if $E$ is obtained we conclude that the states were identical (or equivalently that the photons were indistinguishable), otherwise we assume that the states were different (or equivalently that the first photon was distinguishable from the others). The one-sided error requirement can thus be written as Pr$_i(E)=1$: indistinguishable photons always pass the test. For different input states $\ket{\phi}$ and $\ket{\psi}$, the error probability of the corresponding test is then given by Pr$(E)$, which by Eq.~(\ref{boundpr}) is lower bounded by $\frac{1}{M}+\frac{M-1}{M}|\braket{\phi|\psi}|^2$.

\begin{figure}
\begin{center}
\includegraphics[width=1.7in]{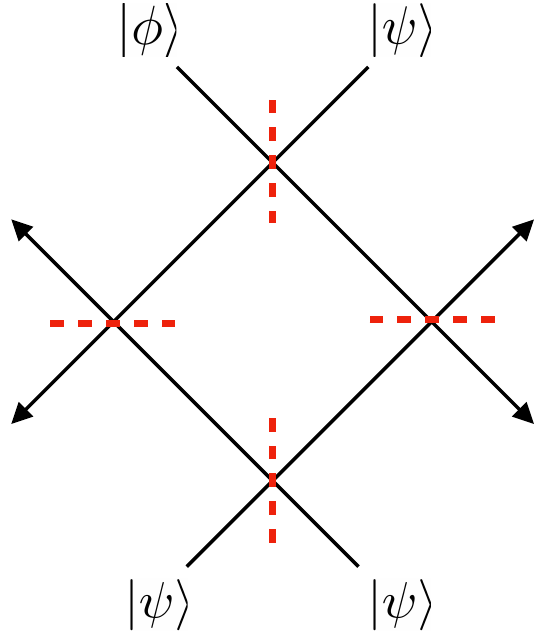}
\caption{Hadamard interferometer with 4 input modes. The dashed red lines represent balanced beam splitters. The input states are one single photon in state $\ket{\phi}$ and three single photons in state $\ket{\psi}$, one in each mode.}
\label{fig:4swap}
\end{center}
\end{figure}

We now study a particular unitary interferometer, when the size $M$ is a power of $2$, namely the Hadamard interferometer~\cite{crespi2015suppression,crespi2016suppression} and show that it provides a practical and simple implementation of the swap test of order $M$. For $M=4$ spatial modes (Fig.~\ref{fig:4swap}), this interferometer is described by the Hadamard-Walsh transform of order $2$:
\begin{equation}
\frac{1}{\sqrt{2}}\begin{pmatrix} H & H \\ H & -H \end{pmatrix} 
\end{equation}
where $H$ is a Hadamard matrix, see Eq.~(\ref{matrixH}). 
In the general case, the Hadamard interferometer of order $M$ is described by the Hadamard-Walsh transform of order $n=\log M$, which is defined by induction:
\begin{equation}
H_{k+1}=\frac{1}{\sqrt{2}}\begin{pmatrix} H_k & H_k \\ H_k & -H_k \end{pmatrix},
\end{equation}
with $H_0=1$ and $H_1=H$. We can now state our main result linking the Hadamard interferometer and the swap test of order $M$.

\begin{theo}  The output statistics of the Hadamard interferometer of order $M$ can be classically post-processed in time $O(M\log M)$ to reproduce those of the swap test of order $M$.
\label{th:Hn}
\end{theo}
\textbf{Proof.} We give hereafter an overview of the proof and refer to the Appendix~\ref{app:proofth} for further details. 

Due to the structure of the Hadamard-Walsh transform, we are able to show that there exists a collection of detection patterns which saturate the bound in Eq.~(\ref{boundpr}) and to characterise this collection. We introduce the $M\times M$ matrix
\begin{equation}
S=(s_{ij})_{0\le i,j\le M-1}=\sqrt{M} H_n,
\end{equation}
thus omitting the normalisation factor. The matrix $S$ only has $+1$ and $-1$ entries. We show that its rows, together with the element-wise multiplication, form a group isomorphic to $\left(\mathbb{Z}/2\mathbb{Z}\right)^n$. We define for all measurement outcomes $D=(d_0,\dots,d_{M-1})$ the function
\begin{equation}
\pi(D)=\sum_{i=0}^{M-1}{\prod_{j=0}^{M-1}{\left(s_{ij}\right)^{d_j}}},
\label{pi}
\end{equation}
and exploit the aforementioned group structure to obtain the following equivalences:
\be
\begin{aligned}
\pi(D)\neq0&\Leftrightarrow\pi(D)=M\\
&\Leftrightarrow \text{Pr}_i(D)\neq0\\
&\Leftrightarrow\text{Pr}_d(D)=\frac{\text{Pr}_i(D)}{M}.
\end{aligned}
\ee
With the first two lines, the condition $\pi(D)= 0$ is directly equivalent to having a detection event $D$ that can only be witnessed in the distinguishable case. In other words, the detection patterns $D$ such that $\pi(D)= 0$ can only occur if $\braket{\phi|\psi}\neq0$. On the other hand, with the third equivalence, the detection patterns $D$ such that $\pi(D)\neq 0$ are those that saturate the bound obtained in Eq.~(\ref{boundpr}). The Hadamard interferometer can thus be used to compare the states $\ket{\phi}$ and $\ket{\psi}$: if the outcome $D$ obtained satisfies $\pi(D)=M$, we conclude that the states were identical, otherwise $\pi(D)=0$ and we conclude that the states were different. We show in particular that the interferometer described by the unitary matrix $H_n$ satisfies
\begin{equation}
\text{Pr}[\pi(D)=M]=\frac{1}{M}+\frac{M-1}{M}\left|\braket{\phi|\psi}\right|^2,
\end{equation}
and
\begin{equation}
\text{Pr}[\pi(D)=0]=1-\text{Pr}[\pi(D)=M],
\end{equation}
for any detection pattern $D$. Hence the identity test using the Hadamard interferometer of order $M$ is a swap test of order $M$. The measurement outcomes $D$ have to be post-processed by computing $\pi(D)$. Using the group structure of the matrix $S$, we show that this can be done in time $O(M\log M)$. 

\qed \\

Note that the group structure invoked in the proof is preserved under permutations, so Theorem~\ref{th:Hn} also applies to the unitary interferometers described by permutations of the Hadamard-Walsh transform.

The conclusion to be drawn from Theorem~\ref{th:Hn} is that as long as a state $\ket{\psi}$ can be encoded using single photons, then one can perform a swap test of order $M$ with respect to the state $\ket{\psi}$ using the Hadamard interferometer of order $M$ and an efficient classical post-processing of the measurement outcomes. The post-processing consists in the following parity test: given the measurement outcome $D=(d_0,\dots,d_{M-1})$, where $d_0+\dots+d_{M-1}= M $, construct the matrix $S_D$ from the matrix $S=\sqrt{M} H_n$ by keeping the $k^{th}$ column only if $d_k$ is odd. If the rows $(1,2,4,\dots,2^{n-1})$ of $S_D$ all have an even number of $-1$, output $0$. Output $1$ otherwise. This means that the post-processing only requires the parity of the photon number in each output mode.

In particular, the photon number resolving detectors can be replaced by detecting the parity of the number of photons in each output mode. Detecting this parity can for example be achieved with microwave technology~\cite{haroche2007measuring,vlastakis2013deterministically,sun2014tracking}. Also only $M-1$ detectors are necessary, since the parity of the number of photon in the remaining mode can be deduced from the parities of the other modes, given that the total number of photons is $M$.

\begin{figure}
\begin{center}
\includegraphics[width=3.3in]{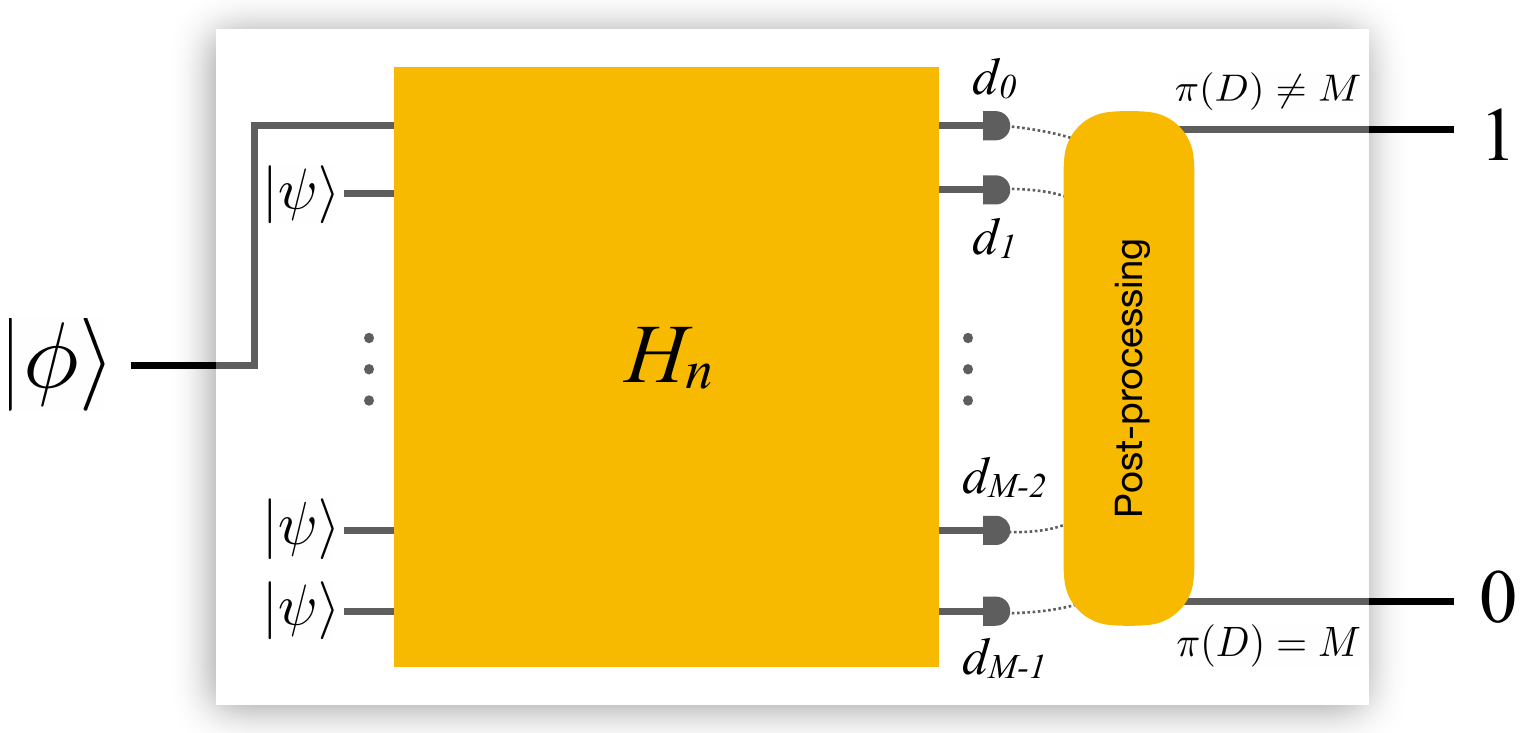}
\caption{The Hadamard interferometer of order $M$ used as a programmable projective measurement device. A single photon in the state $\ket{\phi}$ goes through a linear interferometer along with $M-1$ indistinguishable single photons in the state $\ket{\psi}$. The parity of the number of photons in each output mode is measured and efficiently post-processed, such that the device outputs $0$ with probability $\frac{1}{M}+\frac{M-1}{M}|\braket{\phi|\psi}|^2$ and $1$ with probability $\frac{M-1}{M}(1-|\braket{\phi|\psi}|^2)$.}
\label{fig:setup}
\end{center}
\end{figure}

Using the argument developed in the proof of Theorem~\ref{th:circuit}, by considering the $M-1$ photons and the interferometer as a black box (Fig.~\ref{fig:setup}) whose outcomes are post-processed as described above, we also deduce the following result from Theorem~\ref{th:Hn}:
\begin{coro} The Hadamard interferometer of order $M$ can be used to perform a projective measurement with error $\frac{1}{M}$, using a classical post-processing of its measurement outcomes that takes time $O(M\log M)$.
\label{co:2}
\end{coro}
Interestingly, the unitary interferometers described by the Hadamard-Walsh transform and its permutations are not the only unitary interferometers which can reproduce the statistics of a swap test with efficient post-processing, and indeed we present a generalisation in Sec.~\ref{sec:general}. However, it is the simplicity of the Hadamard interferometer in terms of experimental implementation that motivates our interest towards this interferometer. In particular, this interferometer can be simply implemented with a few balanced beam splitters. A result by Reck \textit{et al.}~\cite{reck1994experimental} states that any $M\times M$ unitary interferometer can be implemented using phase shifters and at most $\frac{M(M-1)}{2}$ beam splitters, possibly unbalanced. For the Hadamard interferometer, only $\frac{M\log M}{2}$ balanced beam splitters are needed and no phase shifters. The proof of this statement is based on a simple induction detailed in Appendix~\ref{app:beam}.
%


\section{Group generalisation for any value of the size parameter $M$}
\label{sec:general}

The Hadamard interferometer requires the size parameter $M$ to be a power of $2$. This requirement can be relaxed, possibly raising the experimental requirements at the same time. Indeed, for any value of $M$, one can associate to any abelian group of order $M$ an interferometer of size $M$ which has the desired statistics. This is the object of the following result that uses the invariant factor decomposition of an abelian group:
\begin{theo} Let $G$ be an abelian group of order $M$. Then there exists $N\in\mathbb{N}^*$ and $a_1,\dots,a_N\in\mathbb{N}^*$, where $a_i|a_{i+1}$ for $i\in\{1,\dots,N-1\}$ and $a_1\dots a_N=M$, such that the interferometer described by the $M\times M$ unitary matrix
\begin{equation}
U_G=\frac{1}{\sqrt{M}}F_{a_1}\otimes\dots\otimes F_{a_N},
\end{equation}
where $F_a=(e^{\frac{2i\pi}{a}kl})_{0\le k,l\le a-1}$ is the Quantum Fourier Transform (QFT) of order $a$ for all $a\in\mathbb{N}^*$, can perform a $\frac{1}{M}$-approximate projective measurement with a post-processing of its measurement outcomes that takes time at most $M\cdot N$. The rows of $F_G=\sqrt{M}U_G$ together with the element-wise multiplication form a group isomorphic to $G$.
\label{th:group}
\end{theo}
\textbf{Proof.} We use the notations of the Theorem. The invariant factor decomposition of $G$ gives
\begin{equation}
G\simeq \left(\mathbb{Z}/a_1\mathbb{Z}\right)\otimes\dots\otimes\left(\mathbb{Z}/a_N\mathbb{Z}\right),
\end{equation}
where $N\in\mathbb{N}^*$ and $a_1,\dots,a_N\in\mathbb{N}^*$ are unique, satisfying $a_i|a_{i+1}$ for $i\in\{1,\dots,N-1\}$ and $a_1\dots a_N=M$. Given that the rows of $F_a$ together with the element-wise multiplication form a group isomorphic to $\left(\mathbb{Z}/a\mathbb{Z}\right)$ for all $a\in\mathbb{N}^*$, the rows of $F_G=(f_{ij})_{0\le i,j\le M-1}=\sqrt{M}U_G$ together with the element-wise multiplication form a group isomorphic to $G$.

Since the group structure was the only argument invoked in the proof of Theorem~\ref{th:Hn}, the same conclusion can be drawn here, by following the same argument:
\begin{equation}
\text{Pr}[\pi(D)=M]=\frac{1}{M}+\frac{M-1}{M}\left|\braket{\phi|\psi}\right|^2,
\end{equation}
where
\begin{equation}
\pi(D)=\sum_{i=0}^{M-1}{\prod_{j=0}^{M-1}{\left(f_{ij}\right)^{d_j}}}.
\end{equation}
The group $G$ is finitely generated by $N$ elements, so $N$ rows of $F_G$ are sufficient to generate all its rows by element-wise multiplication. The condition $\pi(D)=M$ can thus be checked in time at most $M\cdot N$. 

\qed\\

In particular, for $G\simeq(\mathbb{Z}/M\mathbb{Z})$, the corresponding interferometer is described by the (normalised) QFT of order $M$, while for $G\simeq(\mathbb{Z}/2\mathbb{Z})^n$, we retrieve Theorem~\ref{th:Hn} and the Hadamard interferometer.


\section{Conclusion and discussion}
\label{sec:concl}

We have presented an optimal scheme for a programmable projective measurement device, and a linear optical implementation, the Hadamard interferometer, which is straightforward and efficient. 
This could for example be used to design a photonic circuit which would act as a universal projective measurement device for a broad range of potential applications from quantum information and cryptography to tests of contextuality.

The Hadamard interferometer is easily implementable, but this comes at the cost that we are detecting all modes, i.e. that there is no quantum output unlike for the swap circuit of order $M$. However, for most applications, it is only the classical output statistics of the circuit that matters, as it is the case e.g. for quantum state identity testing.

Our scheme can also be interpreted as an optimal swap test when one has a single copy of one state, and $M-1$ of the other. Given the breadth of applications of the swap test for entanglement testing~\cite{mintert2005concurrence,walborn2006experimental,harrow2013testing}, communications~\cite{buhrman2001quantum,de2004one,kumar2017efficient}, quantum machine learning~\cite{ekert2002direct,lloyd2014quantum} e.t.c., one can anticipate our result will have applications also in these domains.

We have chosen to phrase the problem in terms of $M-1$ copies of the state $|\psi\rangle$. In principle we could have chosen any other encoding of the quantum input into $M-1$ registers. 
The reason for our choice is twofold. Firstly it is part of the envisaged problem setting - we imagine a device producing states encoding our measurement, for the example these could be the output of a computation. Secondly we do so in order to separate as much as possible the resource of $M-1$ program systems and the process of translating them into a measurement. In particular if one had any other encoding, for example into some entangled states, this encoding process could be incorporated into the circuit representing the generic measurement apparatus. In this sense the most quantum information that can be contained about the state $|\psi\rangle$ in $M-1$ systems is $M-1$ copies of the state $|\psi\rangle$ - anything more can be done afterwards. See for example \cite{brazier2005probabilistic} for a similar discussion in the case of programmable quantum computation of $U(1)$ rotations.

This result also gives rise to a natural interpretation of the notion of projective measurement in quantum mechanics, as a comparison between one state and several copies of another state using an interferometer: in the macroscopic limit, when many copies of a reference eigenstate are available, we retrieve a macroscopic classically programmable quantum measurement set up.

For completeness, it could be interesting to characterise the full class of interferometers that are optimal for state identity testing under the one-sided error requirement, as we only gave a broad class of such interferometers using a group construction. We conjecture that the Hadamard interferometer will remain the simplest to implement among this class of optimal schemes. It would be also interesting to consider the influence of real experimental conditions, as our scheme assumes that the input states are pure. The one-sided error requirement is also a challenge experimentally, as any interferometer would suffer from the effects of imperfection and noise. We leave these analyses open for future work.


\section{Acknowledgements}

We kindly acknowledge F. Grosshans and A. Olivo for interesting and inspiring discussions. This work has been supported in part by the European Union's H2020 Programme under grant agreement number ERC-669891, by the European Research Council Starting Grant QUSCO and by the ANR COMB project.


\appendix

\section{Proof of optimality}
\label{app:opti}

An identity test on a Hilbert space $\mathcal{H}$ is a binary test which can be written as a positive-operator valued measure $\{\Pi_0,\Pi_1\}$, with $\Pi_0+\Pi_1=I$. Such a test takes as input a pure tensor product state $\ket{\psi_0\dots \psi_{M-1}}\in\mathcal{H}^{\otimes M}$ and outputs $0$ with probability
\be
P(0)=\text{Tr}[\Pi_0\ket{\psi_0\dots \psi_{M-1}}\bra{\psi_0\dots \psi_{M-1}}], 
\ee
and $1$ with probability
\be
P(1)=1-P(0)=\text{Tr}[\Pi_1\ket{\psi_0\dots \psi_{M-1}}\bra{\psi_0\dots \psi_{M-1}}].
\ee
If the output $0$ is obtained we conclude that we had $\ket{\psi_0}=\dots=\ket{\psi_{M-1}}$, whereas if the output $1$ is obtained we conclude that the states were not all identical. The one-sided error requirement can thus be written as
\be
\forall\ket{\psi},\text{ Tr}[\Pi_1\ket{\psi}\bra{\psi}^{\otimes M}]=0.
\label{oser}
\ee
Following~\cite{harrow2013church}, the symmetric subspace of $\mathcal{H}^{\otimes M}$ can be characterised as
\be
S=\text{span}\{\ket{\psi}^{\otimes M}:\ket{\psi}\in\mathcal{H}\},
\ee
and the orthogonal projector onto this space can be written as
\be
P_S=\frac{1}{M!}\sum_{\sigma\in\mathcal{S}_M}{P_\sigma},
\ee
where for all $\sigma\in\mathcal{S}_M$ and all $\ket{\psi_0\dots\psi_{M-1}}\in\mathcal{H}^{\otimes M}$ we have $P_\sigma\ket{\psi_0\dots\psi_{M-1}}=\ket{\psi_{\sigma(0)}\dots\psi_{\sigma(M-1)}}$. Given the characterisation of the symmetric subspace, the one-sided error requirement in Eq.~(\ref{oser}) implies that the supports of $P_S$ and $\Pi_1$ are disjoint. The support of $P_S$ is thus included in the support of $\Pi_0$, given that $\Pi_0+\Pi_1=I$ and this implies in turn that $\Pi_0\ge P_S$ by positivity of $\Pi_0$. 

The error probability of the identity test under the one-sided error requirement is given by the probability of outputting the result $0$ while the states were not all identical:
\be
\begin{aligned}
P(0)&=\text{Tr}[\Pi_0\ket{\psi_0\dots \psi_{M-1}}\bra{\psi_0\dots \psi_{M-1}}]\\
&\ge\text{Tr}[P_S\ket{\psi_0\dots \psi_{M-1}}\bra{\psi_0\dots \psi_{M-1}}]\\
&\ge\frac{1}{M!}\sum_{\sigma\in\mathcal{S}_M}{\text{Tr}[P_\sigma\ket{\psi_0\dots \psi_{M-1}}\bra{\psi_0\dots \psi_{M-1}}]}\\
&\ge\frac{1}{M!}\sum_{\sigma\in\mathcal{S}_M}{\text{Tr}[\ket{\psi_{\sigma(0)}\dots\psi_{\sigma(M-1)}}\bra{\psi_0\dots \psi_{M-1}}]}\\
&\ge\frac{1}{M!}\sum_{\sigma\in\mathcal{S}_M}{\prod_{k=0}^{M-1}{\braket{\psi_k|\psi_{\sigma(k)}}}},
\end{aligned}
\ee 
where in the third line we used the expression of the orthogonal projector $P_S$ onto the symmetric subspace.


\section{Statistics of an interferometer}
\label{app:stat}

Recall that we consider optical unitary interferometers of size $M$ which take as input one single photon in a quantum state $\ket{\phi}$ and $M-1$ indistinguishable single photons in a state $\ket{\psi}$, one in each spatial mode, indexed from $0$ to $M-1$. The output modes are measured using photon number detection. A measurement outcome thus has the form $D=(d_0,\dots,d_{M-1})$, with \mbox{$d_0+\dots+d_{M-1}=M$}.

The permanent of an $M\times M$ matrix $T=(t_{ij})_{0\le i,j\le M-1}$ is defined by
\begin{equation}
\text{Per}(T)=\sum_{\sigma\in \mathcal{S}_M}{\prod_{k=0}^{M-1}{t_{k\sigma(k)}}},
\end{equation}
where $\mathcal{S}_M$ is the symmetric group over $\{0,\dots,M-1\}$. We now compute Pr$_i(D)$ and Pr$_d(D)$ for all detection patterns $D$.

In the indistinguishable case, $M$ indistinguishable photons, one in each mode, are sent through a linear optical network described by an $M\times M$ unitary matrix $U=(u_{ij})_{0\le i,j\le M-1}$. The probability of a detection event $D$ can be computed (see, e.g,~\cite{Aaronson2013}) as
\begin{equation}
\text{Pr}_i(D)=\frac{|\text{Per}(U_D)|^2}{D!},
\label{pru}
\end{equation}
where $D!=d_0!\dots d_{M-1}!$ and where $U_D$ is the matrix obtained from $U$ by repeating $d_k$ times the $k^{th}$ column for $k\in\{0,\dots,M-1\}$. 

In the distinguishable case, $M-1$ indistinguishable photons are sent in modes $1,\dots,M-1$ through a linear optical network described by an $M\times M$ unitary matrix $U=(u_{ij})_{0\le i,j\le M-1}$, along with one additional photon in the $0^{th}$ mode in an orthogonal state. Since it is fully distinguishable from the others, the additional photon behaves independently, hence the probability of detecting the photon number pattern $D$ for one distinguishable photon and $M-1$ indistinguishable photons in input is
\begin{equation}
\text{Pr}_d(D)=\sum _{ \substack{k=0\\d_k\neq0} }^{ M-1 }{ \text{Pr}_i(D-1_k)\cdot\text{Pr}_i(1_k) }. 
\end{equation}
This last expression formalises the fact that the $M-1$ indistinguishable photons give a detection pattern \mbox{$D-1_k$} which, completed by the additional distinguishable photon in the $k^{th}$ output mode, forms the pattern $D$. Developing this expression with Eq.~(\ref{pru}) yields
\begin{equation}
\text{Pr}_d(D)=\frac { 1 }{ D! } \sum _{ \substack{k=0\\d_k\neq0} }^{ M-1 }{ d_{ k }|u_{ 0k }\text{Per}(U_{ 0,D-1_{ k } })|^{ 2 } } 
\label{prd}
\end{equation}
where $U_{0,D-1_{k}}$ is the matrix obtained from $U$ by removing the $0^{th}$ row, then by repeating $d_l$ times the $l^{th}$ column for $l\neq k$ and by repeating $d_k-1$ times the $k^{th}$ column.

In order to obtain more readable expressions, we define for all $k\in\{0,\dots,M-1\}$ and for any detection pattern $D$,
\begin{equation}
p_k(D)=\begin{cases}  \frac{u_{0k}\text{Per}(U_{0,D-1_k})}{\sqrt{D!}} \text{ if } d_k\neq0, \\ 0\text{ otherwise.} \end{cases}
\end{equation}
Using the Laplace expansion of the permanent, the previous equations~(\ref{pru},~\ref{prd}) rewrite
\begin{equation}
\text{Pr}_i(D)=\left|\sum_{k=0}^{M-1}{d_kp_k(D)}\right|^2,
\label{pru2}
\end{equation}
and
\begin{equation}
\text{Pr}_d(D)=\sum_{k=0}^{M-1}{d_k|p_k(D)|^2}.
\label{prd2}
\end{equation}
Since $\sum_{k=0}^{M-1}{d_k}=M$, we obtain, using Cauchy-Schwarz inequality with the complex vectors $\left\{\sqrt{d_k}\right\}_{0\le k\le M-1}$ and $\left\{\sqrt{d_k}p_k(D)\right\}_{0\le k\le M-1}$,
\begin{equation}
\text{Pr}_d(D)\ge\frac{\text{Pr}_i(D)}{M},
\label{general}
\end{equation}
for any detection pattern $D$.


\section{Proof of Theorem~\ref{th:Hn}}
\label{app:proofth}

Let us define
\begin{equation}
S=(s_{ij})_{0\le i,j\le M-1}=\sqrt{M} H_n,
\end{equation}
thus omitting the normalisation factor. We have
\begin{equation}
S=\underbrace { \sqrt{2}H\otimes \dots \otimes \sqrt{2}H }_{ n\text{ }times },
\end{equation}
where $H$ is a Hadamard matrix. The rows of $\sqrt{2}H$, together with the element-wise multiplication, form a group isomorphic to $\mathbb{Z}/2\mathbb{Z}$, thus the rows of $S$ together with the element-wise multiplication form a group isomorphic to $\left(\mathbb{Z}/2\mathbb{Z}\right)^n$. As a consequence, multiplying element-wise all the rows of $S$ by its $i^{th}$ row for a given $i$ amounts to permuting the rows of $S$. Let $D=(d_0,\dots,d_{M-1})$ and $k\in\{0,\dots,M-1\}$ such that $d_k\neq0$. Let also $S_{D-1_k}$ be the matrix obtained from $S$ by repeating $d_l$ times the $l^{th}$ column for $l\neq k$ and $d_k-1$ the $k^{th}$ column. For all $i\in\{0,\dots,M-1\}$, one can obtain the matrix $S_{0,D-1_k}$ (with the $0^{th}$ row removed) from the matrix $S_{i,D-1_k}$  (with the $i^{th}$ row removed) by multiplying element-wise all rows by the $i^{th}$ row and permuting the rows. Since the permanent is invariant by row permutation we obtain, for all $i\in\{0,\dots,M-1\}$ and all $k\in\{0,\dots,M-1\}$ such that $d_k\neq0$,
\begin{equation}
\text{Per}(S_{i,D-1_k})=\epsilon_{ik}(D)\text{Per}(S_{0,D-1_k}),
\label{pmper}
\end{equation}
where $\epsilon_{ik}(D)=s_{ik}\prod_{j=0}^{M-1}{\left(s_{ij}\right)^{d_j}}$. Finally, we use the Laplace row expansion formula for the permanent of $S_D$ to obtain, for all $D=(d_0,\dots,d_{M-1})$ and all \mbox{$k\in\{0,\dots,M-1\}$} such that $d_k\neq0$,
\begin{equation}
\begin{aligned}
\text{Per}(S_D)&=\sum_{i=0}^{M-1}{s_{ik}\text{Per}(S_{i,D-1_k})}\\
&=\left(\sum_{i=0}^{M-1}{s_{ik}\epsilon_{ik}(D)}\right)\text{Per}(S_{0,D-1_k})\\
&=\left(\sum_{i=0}^{M-1}{\prod_{j=0}^{M-1}{\left(s_{ij}\right)^{d_j}}}\right)\text{Per}(S_{0,D-1_k})\\
&=\pi(D)\text{Per}(S_{0,D-1_k}),
\end{aligned}
\label{perpi}
\end{equation}
where we used Eq.~(\ref{pmper}) in the second line. With the general expressions of Pr$_i(D)$~(\ref{pru}) and Pr$_d(D)$~(\ref{prd}), this equation implies 
\begin{equation}
M\text{Pr}_i(D)=\pi(D)^2\text{Pr}_d(D).
\label{prupiprd}
\end{equation}
With the Laplace column expansion formula for the permanent of $S_D$ and the last line of Eq.~(\ref{perpi}), we also obtain
\begin{equation}
M^2\text{Pr}_i(D)=\pi(D)^2\text{Pr}_i(D).
\label{pruprd}
\end{equation}
In particular, combining Eqs.~(\ref{prupiprd},\ref{pruprd}),
\begin{equation}
M^2\pi(D)^2\text{Pr}_d(D)=\pi(D)^4\text{Pr}_d(D).
\end{equation}
Now Pr$_d(D)$ is non-zero for all $D$, since by Eq.~(\ref{prd}) it is a sum of moduli squared of permanents of $(2^n-1)\times(2^n-1)$ matrices, which in turn cannot vanish by a result of~\cite{simion1983+}. Hence the previous equation rewrites
\begin{equation}
M\pi(D)=\pi(D)^2.
\label{piM0}
\end{equation}
As a consequence, $\pi(D)=M$ or $\pi(D)=0$ for all $D$. Combining Eqs.~(\ref{prupiprd},\ref{piM0}) we obtain
\begin{equation}
\begin{aligned}
\pi(D)\neq0&\Leftrightarrow\pi(D)=M\\
&\Leftrightarrow \text{Pr}_i(D)\neq0\\
&\Leftrightarrow\text{Pr}_d(D)=\frac{\text{Pr}_i(D)}{M},
\end{aligned}
\label{impl}
\end{equation}
and thus 
\begin{equation}
\begin{aligned}
\text{Pr}_i[\pi(D)=M]&=\sum_{\pi(D)=M}{\text{Pr}_i(D)}\\
&=\sum_{\text{Pr}_i(D)\neq0}{\text{Pr}_i(D)}\\
&=1.
\end{aligned}
\label{pru2}
\end{equation}
We also obtain
\begin{equation}
\begin{aligned}
\text{Pr}_d[\pi(D)=M]&=\sum_{\pi(D)=M}{\text{Pr}_d(D)}\\
&=\frac{1}{M}\sum_{\pi(D)=M}{\text{Pr}_i(D)}\\
&=\frac{1}{M}.
\end{aligned}
\label{prd2}
\end{equation}
We finally conclude by combining Eqs.~(\ref{pru2},\ref{prd2}) and Eq.~(\ref{partialud}):
\begin{equation}
\begin{aligned}
\text{Pr}[\pi(D)=M]&=\sum_{\pi(D)=M}{\text{Pr}(D)}\\
&=\frac{1}{M}+\frac{M-1}{M}\left|\braket{\phi|\psi}\right|^2.
\end{aligned}
\end{equation}

The post-processing mentioned in the main text, i.e. computing $\pi(D)$, can be done efficiently in time $O(M\log M)$ for any detection pattern $D=(d_0,\dots,d_{M-1})$. Indeed, let $S_D$ be the $M\times M$ matrix obtained from $S$ by repeating $d_k$ times the $k^{th}$ column for \mbox{$k\in\{0,\dots,M-1\}$}. The expression $\pi(D)$ in Eq.~(\ref{pi}) is the sum of the product of the elements of each row of $S_D$. Since the entries of the matrix $S$ are only $+1$ and $-1$, $\pi(D)=M$ if and only if the number of $-1$ on the rows of $S_D$ is even for all rows. The condition $\pi(D)=M$ can thus be written as a system of $M$ linear equations modulo $2$. Since $\left(\mathbb{Z}/2\mathbb{Z}\right)^n$ is finitely generated by $n$ elements, the $M$ rows of $S_D$ can be generated with at most $n$ rows using element-wise multiplication, for any measurement outcome $D$. Hence, computing the parity of the number of $-1$ on each row of $S_D$, which is equivalent to testing $\pi(D)=M$, can be done by computing at most $n=\log M$ parity equations, with a number of terms in each equation which is at \mbox{most $M$.} 

A simple induction shows that a possible choice for the rows whose parity has to be tested is the rows with index $2^k$ for $k\in\{0,\dots,n-1\}$ (the rows of the matrix being indexed from $0$ to $M-1$).\\ \\


\section{The Hadamard interferometer can be implemented with a few balanced beam splitters}
\label{app:beam}

Let $I_k$ be the $k\times k$ identity matrix for all $k$. The size $M$ is a power of $2$, with $n=\log M$. We prove by induction over $n$ that, there exist $P_0(n),\dots,P_{n-1}(n)$ permutation matrices of order $M/2$, such that
\begin{equation}
H_n=\prod _{ k=0 }^{ n-1 }{ { P }_{ k }(n)\left( { I }_{ M/2 }\otimes { H } \right) { { P } }_{ k }(n)^{ T } }.
\label{hyprec}
\end{equation}
Since multiplying matrices is equivalent to setting up experimental devices in sequence, and given that $H$ is the matrix describing a balanced beam splitter, Eq.~(\ref{hyprec}) implies the result we want to prove.

For $n=1$, we have $M=2$ and Eq.~(\ref{hyprec}) is true with $P_0(1)=I_1$.  For brevity, we define for all $k$
\begin{equation}
H^{(k)} ={ I }_{ k }{ \otimes} H.
\end{equation}
Assuming that Eq.~(\ref{hyprec}) is true for $n$, we use the recursive definition of the Hadamard-Walsh transform
\begin{equation}
H_{n+1} = H\otimes H_n,
\end{equation}
along with properties of the tensor product of matrices in order to obtain
\begin{equation}
\begin{aligned}
&H_{n+1} = \left( { H }_{ n }\otimes { I }_{ 2 } \right) H^{(M)} 
= Q\left( { I }_{ 2 }\otimes { H }_{ n } \right) { Q }^{ T }H^{(M)} \\ 
&= Q\left[ { I }_{ 2 }\otimes \prod _{ k=0 }^{ n-1 }{ { P }_{ k }(n)H^{(M/2)} { { P } }_{ k }(n )^{ T } }  \right] { Q }^{ T }H^{(M)}  \\ 
&= Q\left[ \prod _{ k=0 }^{ n-1 }{ \left( { I }_{ 2 }\otimes { P }_{ k }(n) \right) H^{(M)} \left( { I }_{ 2 }\otimes { { P } }_{ k }(n)^{ T } \right)  }  \right] { Q }^{ T }H^{(M)} \\ 
&= \prod _{ k=0 }^{ n-1 }{ \left[ Q\left( { I }_{ 2 }\otimes { P }_{ k }(n) \right)  \right] H^{(M)} { \left[ { Q }\left( { I }_{ 2 }\otimes { { P } }_{ k }(n) \right)  \right]  }^{ T } }H^{(M)},
\end{aligned}
\end{equation}
where $Q$ is a permutation matrix of order $M$ and where in the third line we have used Eq.~(\ref{hyprec}). Setting $P_k(n+1)=Q\left( { I }_{ 2 }\otimes { P }_{ k }(n)\right)$ for $k\in\{0,\dots,n-1\}$ and $P_{n}(n+1)=I_{M}$ proves Eq.~(\ref{hyprec}) for $n+1$, since these matrices are permutation matrices of order $M$. This completes the induction and the proof of the result.


\bibliographystyle{apsrev}
\bibliography{bibliography}

\end{document}